\documentclass[conference]{IEEEtran}
\IEEEoverridecommandlockouts
\usepackage{cite}
\usepackage{amsmath,amssymb,amsfonts}
\usepackage{algorithmic}
\usepackage{graphicx}
\usepackage{textcomp}
\usepackage{xcolor}
\usepackage{comment}

\usepackage{fancyhdr}

\usepackage{caption}
\usepackage{siunitx}
\usepackage{booktabs}
\usepackage{tabularx}
\newcolumntype{Y}{>{\centering\arraybackslash}X}

\def\BibTeX{{\rm B\kern-.05em{\sc i\kern-.025em b}\kern-.08em
    T\kern-.1667em\lower.7ex\hbox{E}\kern-.125emX}}
\begin{document}

\title{Wise in Vaccine Allocation\\

}

\makeatletter 
\newcommand{\linebreakand}{%
  \end{@IEEEauthorhalign}
  \hfill\mbox{}\par
  \mbox{}\hfill\begin{@IEEEauthorhalign}
}
\makeatother 

\author{
\IEEEauthorblockN{Baiqiao Yin\IEEEauthorrefmark{2}}
\IEEEauthorblockA{\textit{School of Intelligent Systems Engineering} \\
\textit{Sun Yat-sen University}\\
Shenzhen, China \\
yinbq@mail2.sysu.edu.cn}

\and
\IEEEauthorblockN{Jiaqing Yuan}
\IEEEauthorblockA{\textit{School of Economics and Management} \\
\textit{Dalian University of Technology}\\
Dalian, China \\
yjquii@163.com}
\and
\IEEEauthorblockN{Weichen Lv}
\IEEEauthorblockA{\textit{School of Intelligent Systems Engineering} \\
\textit{Sun Yat-sen University}\\
Shenzhen, China \\
lvwch@mail2.sysu.edu.cn}

\linebreakand 

\IEEEauthorblockN{Guian Fang\thanks{\IEEEauthorrefmark{1}Corresponding author}\IEEEauthorrefmark{1}}
\IEEEauthorblockA{\textit{School of Intelligent Systems Engineering} \\
\textit{Sun Yat-sen University}\\
Shenzhen, China \\
fanggan@mail2.sysu.edu.cn}
\and

\IEEEauthorblockN{Jiehui Huang\thanks{\IEEEauthorrefmark{2}Co-first author}\IEEEauthorrefmark{2}}
\IEEEauthorblockA{\textit{School of Intelligent Systems Engineering} \\
\textit{Sun Yat-sen University}\\
Shenzhen, China \\
huangjh336@mail2.sysu.edu.cn}

}

\maketitle

\thispagestyle{fancy}
\fancyhead{} 
\fancyfoot{}
\lhead{IEEE 6th International Conference on Universal Village · UV2022 · Session TS11-D-Vac-8} 
\lfoot{\fontsize{8}{10} \selectfont 978-1-6654-7477-1/22/\$31.00 \copyright2022 IEEE} 
\rhead{}
\cfoot{} 
\rfoot{}

\begin{abstract}
In this paper, the machine learning method and mathematical model are used to predict the number of future vaccinations, and the problem of how to distribute vaccines to central  hospitals, community  hospitals  and  health  centers  is solved\cite{Lyu_Han_Luli_2021}, \cite{Matrajt_Eaton_Leung_Dimitrov_Schiffer_Swan_Janes_2021}. In the context of the growing importance of vaccination, we need to rationalize the distribution of vaccines to central hospitals, community hospitals and health centers, taking into account  the  need  and  cost  of vaccination. First,  in  order  to predict the national daily vaccination figures for the next three months,  we  consulted  relevant  website  data  to  obtain  the vaccination figures for each day since the vaccination began in March 2021, and made the forecast for the next three months through the time  series prediction  method LSTM\cite{1997Long}, \cite{Vlachas2018Data}. Combined with the increment of the number of daily vaccinations as the label value, the final prediction results were obtained. Second, we   first   collected   data   and   analyzed   and   processed   the characteristics. Through collinearity analysis\cite{Wang_Nayak_Guttery_Zhang_Zhang_2020}, we found that the number of residents and the number of medical personnel had strong collinearity, and the logarithm of the number of residents was calculated with log10. Then AHP\cite{SAATY1987161} was used to analyze the impact   of   the   number   of   nearby   residents,  convenient transportation, number of medical personnel, vaccine storage and transportation costs on vaccine distribution, and CR index was  used  to  evaluate  our  model. The  third  question  is  to substitute the collected data of the two regions into the model of the previous question, and we subtract 10\% number of nearby residents from the index of central hospitals as a penalty for crowd  gathering. Got  central  hospitals, community  hospitals, and health centers vaccine distribution ratio: Hangzhou Gongshu District 4.8:3.3:1.9; Harbin Daoli District 3.6:4.7:1.7\cite{Soares_Rocha_Moniz_Gama_Laires_Pedro_Dias_Leite_Nunes_2021}. Fourth, in combination with our model and conclusions, we provide an adequate explanation for vaccine distribution.
\end{abstract}

\hfill  


\section{Introduction}
\subsection{Promblem Background}
The epidemic of the past three years has brought great disasters to mankind and changed the way people live. And the Type V vaccine has saved tens of millions of lives worldwide, and remains the most  important  way  to  control  COVID-19,  and  move  the pandemic to the next stage.

As epidemic control policy in China changes, promoting vaccination is even more important, especially among the elderly. To make it easier for the public to vaccinate, we will add   additional   vaccination   sites   in   central   hospitals, community hospitals and health centers. However, due to the cost of vaccine transportation and storage, we must consider how to distribute the vaccine to central hospitals, community hospitals and health centers.

\subsection{Promblem Restatement}
\begin{enumerate}
    \item[$\bullet$]  Problem \uppercase\expandafter{\romannumeral1}: predict   and  visualize  the  national   daily  vaccination numbers for the next three months.
    \item[$\bullet$]  Problem \uppercase\expandafter{\romannumeral2}: considering  the  number  of  nearby  residents,  the convenience   of  transportation, the   number   of  medical personnel, vaccine  storage  and  transportation  costs, while avoiding excessive gathering of people during vaccination, design  the  vaccine  distribution  plan  for  central  hospitals, community hospitals and health centers.
    \item[$\bullet$]  Problem \uppercase\expandafter{\romannumeral3}: illustrated  Gongshu  District  of  Hangzhou  and  Daoli District  of Harbin  as  examples, calculate  the  number  or proportion of vaccines issued by central hospitals, community hospitals and health hospitals in the two districts.
    \item[$\bullet$]  Write down the instructions on vaccine distribution.
\end{enumerate}

\subsection{Explanations of Symbols}
Explanations of symbols are shown in Table \uppercase\expandafter{\romannumeral1}.
\begin{table}[!ht]
    \centering
    \caption{Explanations of Symbols}
    \begin{tabular}{cc}
    \hline
        \textbf{Symbol} & \textbf{Meaning}  \\ \hline
        \textbf{NoR} & number of nearby residents \\ 
        \textbf{TC} & transportation convenience  \\ 
        \textbf{NoS} & number of medical staff  \\ 
        \textbf{Cost} & vaccine storage and transportation costs \\ 
        \textbf{CenH} & central hospitals  \\ 
        \textbf{ComH} & community hospitals  \\ 
        \textbf{HC} & health centers \\ \hline
    \end{tabular}
\end{table}

\section{Model and Solution of Problem 1}
\subsection{Time Series Prediction about the Number of Vaccination}
In order to predict the national daily vaccination figures in the next three months, first, we checked the relevant website data, investigated  the  number   of  everyday  vaccination population since vaccination in the past two years (March 23, 2021 to December 18, 2022), made statistics, and took the timing changes as our original data set.
\begin{figure}[h]
    \centering
    \includegraphics{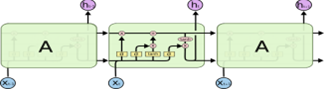}
    \caption{LSTM module diagram}
    \label{fig:1}
\end{figure}

Next, the mining of the time-series relationship of the number of people vaccinated was performed. For this part, we used the classical LSTM model\cite{1997Long}, \cite{Vlachas2018Data} in the deep learning field for temporal prediction. As shown in Fig. \ref{fig:1}, LSTM (Long Short-Term Memory) is a special recurrent neural network, firstly established in the NLP field, which can learn the long-term dependencies in sequence data, mainly to solve the problem of gradient disappearance and  gradient  explosion  in  the  training  process  of  a long sequence. It controls the flow of information by introducing a gate  mechanism,  allowing  the  model  to  process  the  long sequence data better, and to remember the information for longer times. Its main working components have three parts:
\begin{itemize}
    \item Forget gate: acting on the cell state and enabling the information selective forgetting in the cell state
\item Input layer gate: acting on the cell state and selectively recording new information into the cell state
\item Output layer gate: acting on the hidden layer ht and saving the previous information to the hidden layer

\end{itemize}

After that, we directly put the raw data into the model for training  and  prediction, but  the  effect  is  not  as  good  as expected. The model fits the actual label well on the whole, which can be seen in Fig. \ref{fig:2}. But there is a large gap between it and the real value in detail, which is unacceptable to us. After analysis, we believe that the reason for this problem is the few characteristics of the data set, because we want to directly predict the number of people receiving the vaccine, without paying any more details.

In order to solve the problem of the predicted value on the details and the real value differing greatly, we finally choose the daily vaccination number increment as a label value. As can be seen in Fig. \ref{fig:3}, although the after-training model for some odd value is not well fit, but the global effect is considerable, and the error is smaller in detail.
\begin{figure}[h]
    \centering
    \includegraphics[width=0.5\textwidth,height=0.35\textwidth]{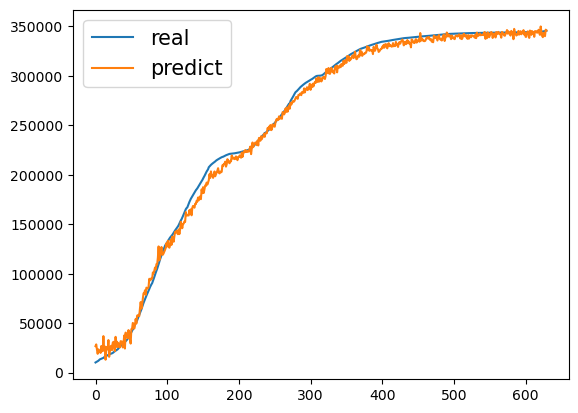}
    \caption{The raw data were substituted into the fit}
    \label{fig:2}
\end{figure}
\begin{figure}[h]
    \centering
    \includegraphics[width=0.5\textwidth,height=0.35
    \textwidth]
    {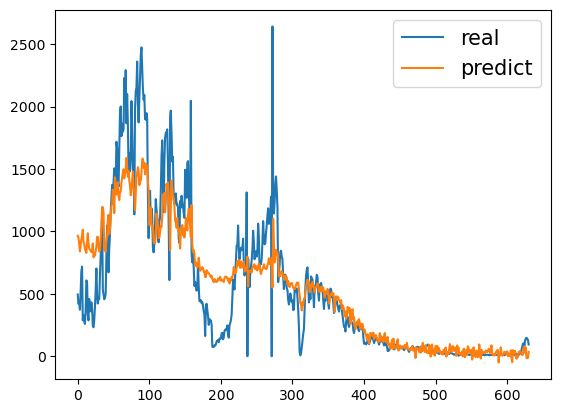}
    \caption{Model after training}
    \label{fig:3}
\end{figure}
\subsection{Visualization of the Prediction Results}
Our final forecast of the number of vaccinated in three months,  as  shown  in  Fig. \ref{fig:4}, will  reach  3467838,700, appearing an increase of more than 10 million from a month ago. The results are fully justified: after nearly two years of publicity and popularization of COVID-19 vaccination, the number  of  COVID-19  citizens  is  already  large, and  the number increased by 13 million in December 2022 compared to November 2022. So we believe that the prediction results are relatively reasonable.
\begin{figure}[h]
    \centering
    \includegraphics[width=0.5\textwidth,height=0.4\textwidth]{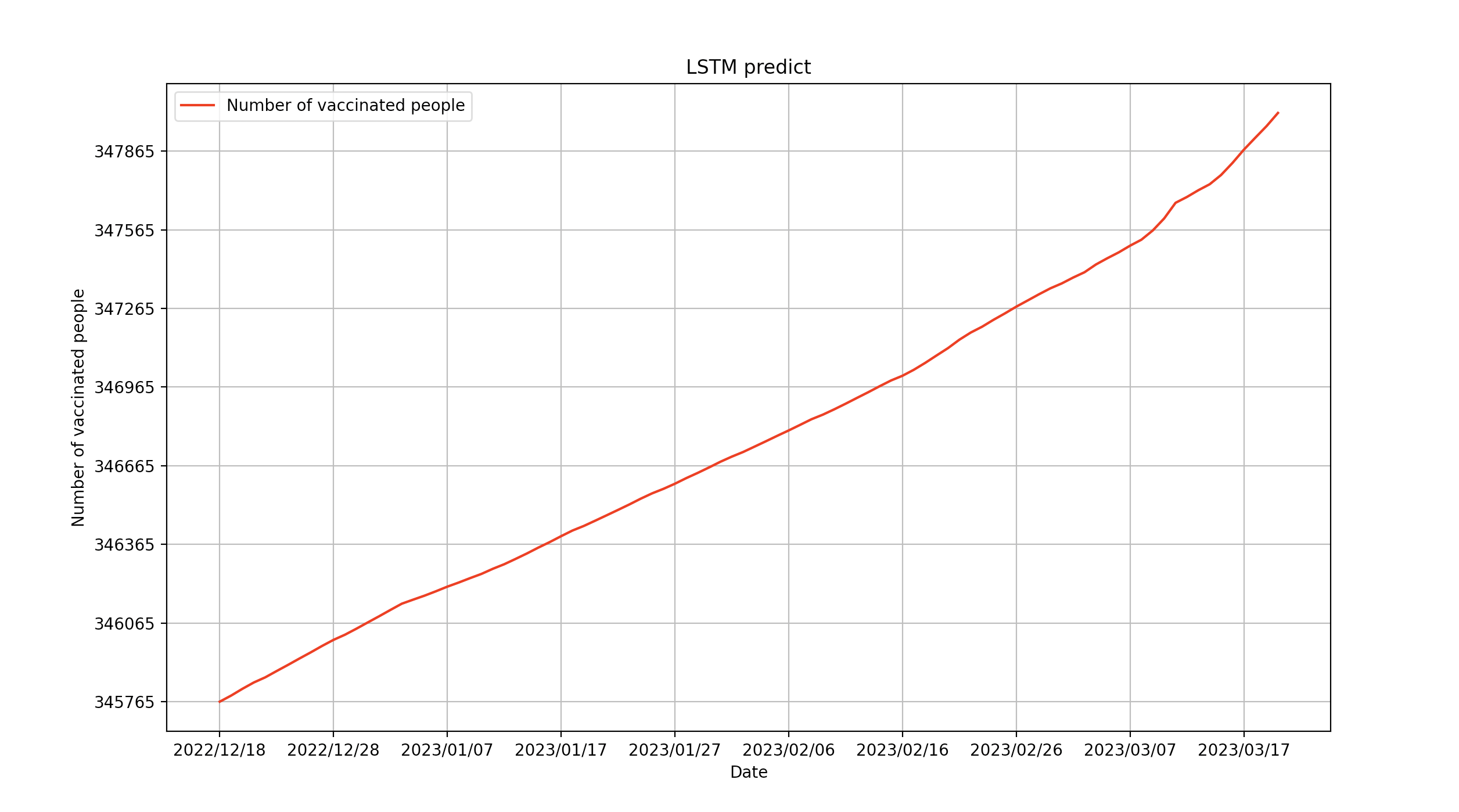}
    \caption{Visualization of the prediction results}
    \label{fig:4}
\end{figure}
\section{Model and Solution of Problem 2}
Ensuring the need for vaccination requires consideration of the number of residents covered by the different health care facilities, the maximum number of health care workers who can provide services, and the accessibility of transportation for these   residents   to   reach   the   health   care   facilities   for vaccination. At the same time, the cost of vaccine storage and transportation  needs  to  be  analyzed  for  different  medical institutions. At  the  same  time, to  avoid  excessive  crowd aggregation during vaccination, it is necessary to restrict the medical institutions that have too many people vaccinated. Therefore, it is necessary to establish a quantitative model for vaccine allocation to central hospitals, community hospitals and  health  centers  considering  the  above   factors. The commonly used comprehensive evaluation methods include AHP, TOPSIS, entropy method, G1, grey correlation analysis, game  theory  combination  weight,  fuzzy  comprehensive evaluation, and Critic comprehensive weighting. We used the AHP+ entropy method to analyze the contribution degree of different factors.
\subsection{Feature Analysis}
\subsubsection{Number of residents and medical personnel}
It occurred to us that because the number of medical staff
in each health care institution may have matched the number of residents  served by each health care institution through development and mobilization over time, this means that there may be strong collinearity and coupling between the number of residents  and  the  number  of medical  staff providing services.

In  order  to  confirm  our  conjecture,  we  collected  the number of residents and medical staff under the jurisdiction of the  district  administrative  units  of  each  city  in  Zhejiang province  and  Heilongjiang  Province. In  addition, in  the process of data collection, we took into account that not all medical institutions in the jurisdiction have the ability and qualification   to   administer   vaccines. Therefore, when collecting data on the medical staff, we did not simply count the number of medical institutions in the jurisdiction, but obtained the information on vaccination sites through the public channel of epidemic prevention information and made statistics.

To analyze the degree of collinearity between the number of residents and the number of medical staff, we performed a t-test for regression coefficient significance, which is a test for independent  variables  to   see  whether  each  independent variable has significant predictive power. The result of t-test was VIF  (Variance  Inflation  Factor)  equal to 33.6. However, in general, for  two  independent  variables, the  value  of VIF should  be  around  5  and  should  not  be  greater  than  10. Therefore, we  conclude  that  there  is  a  strong  collinearity between the number of residents and the number of medical staff.

To mitigate collinearity between the number of residents and the number of health care workers and to account for the diminishing  marginal  effect  of  increasing  the  number  of residents  on  vaccination, we  performed  a  log  base  of  10 operations on the available data on the number of residents to make the number of residents change logarithmically.
\subsubsection{Transportation convenience and vaccine storage and transportation cost}
In terms of transportation convenience, we found that for residents who want to be vaccinated, community hospitals and health centers near their homes are far more attractive than central hospitals. Therefore, we realized that the analysis of transportation convenience is not to analyze the speed of the road network near medical institutions, but the distance of medical  institutions  from  residential  areas, which  can  be expressed as Manhattan distance.

By  analyzing  the  storage  and  transportation  costs  of vaccines and consulting policies and data, we know that the vaccines used in various vaccination sites in our country are uniformly allocated by the whole province, and the storage, distribution, transportation and use are required to  strictly refer to the relevant national laws and regulations, and all vaccines  are  subject  to  24-hour  temperature  monitoring. Therefore, for different central hospitals, community hospitals and health centers in a district, vaccines are all from provincial centers, and  the  difference  in  transportation  costs  due  to distance  is  actually  small. The  cost  difference  is  mainly reflected in vaccine storage. Small-scale hospitals may not have large, centralized cold storage facilities, but with the improvement  of  equipment  and  the  support  of  national epidemic prevention policies, storage will become less of a cost problem.
\subsubsection{Other factors}
Considering   the   need   to   avoid   excessive   crowd aggregation  during  vaccination  and  other  emergencies,  a variable proportional to the number of residents can be set to represent the impact of emergencies and force majeure factors on the selection of vaccinated people.
\subsection{Methods Application}
Analytic hierarchy process (AHP)\cite{SAATY1987161} is a qualitative and quantitative   decision-making   method   to   solve   complex problems with multiple objectives. This method combines quantitative analysis with qualitative analysis, and empirically judges the relative importance of the criteria that measure whether the goals can be achieved, and gives the weight of each criterion of each decision scheme reasonably, then uses the  weight  to  find  out  the  order  of the  advantages  and disadvantages of each scheme. We used AHP to analyze the extent to which the number of nearby residents, accessibility, number of medical staff, vaccine storage, and transportation costs affected vaccine allocation:
\begin{itemize}
    \item Stratify the problems to be analyzed
\item Decompose the problem into different components
\item Condense and combine the factors at different levels to form a multi-level analysis structure model
\item Compare and rank the questions

\end{itemize}
\subsubsection{Establish a hierarchical structure model}
\begin{figure}[h]
    \centering
    \includegraphics[width=0.4\textwidth,height=0.3\textwidth]{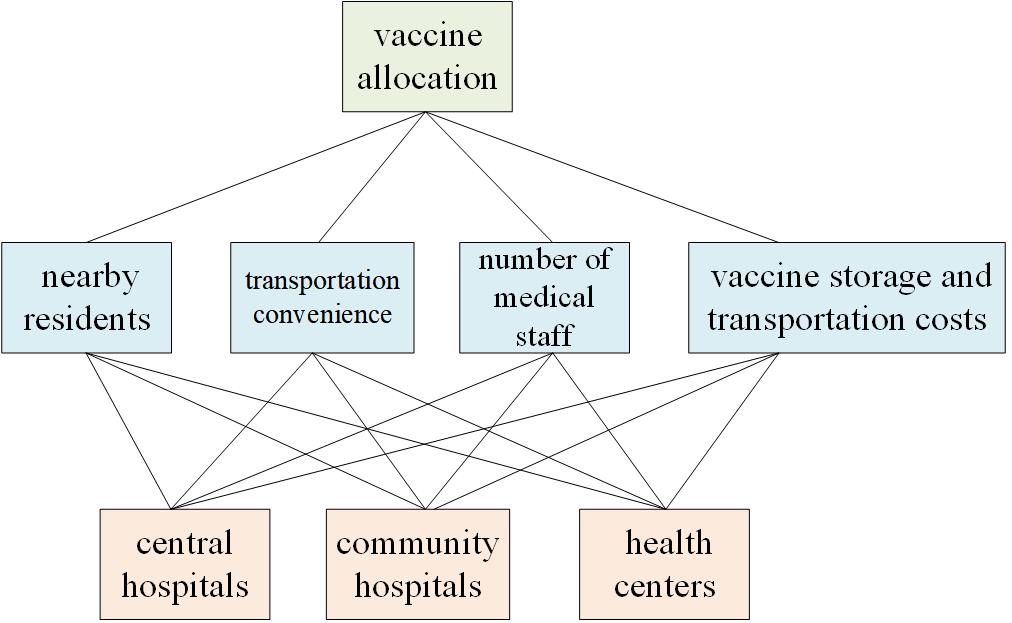}
    \caption{Hierarchy structure chart}
    \label{fig:5}
\end{figure}
The goal, factors (decision criteria) and decision objects are divided into the highest level, the middle level and the lowest level according to their mutual relationship, and the hierarchical structure diagram is drawn in Fig. \ref{fig:5}.
\begin{itemize}
    \item Top level: the purpose of the decision, the problem to be solved, namely vaccine allocation
\item The  lowest  level:  the  alternatives  at  the  time  of the decision,     namely     central     hospitals, community hospitals and health centers
\item Middle layer: The factors considered, the criteria for decision     making,     namely     nearby     residents, transportation, convenience, number of medical staff, vaccine storage and transportation costs

\end{itemize}
\subsubsection{Construct judgment matrix}
The method of constructing a judgment matrix in AHP is the  consensus  matrix  method,  that  is,  all  factors  are  not compared together, but compared with each other. In this case, the relative scale is used to reduce the difficulty of comparing different  factors  with  each  other  as  much  as  possible  to improve the accuracy. The labeling method for the judgment is shown in Table \uppercase\expandafter{\romannumeral2}.
\begin{table}[!ht]
    \centering
    \caption{Method of Marking}
    \begin{tabular}{cc}
    \hline
        \textbf{Scaling} & \textbf{Meaning} \\ \hline
        \specialrule{0em}{1pt}{3pt}
        \textbf{1} & Indicates that compared with two factors,  \\ & they have the same importance \\ 
        \specialrule{0em}{3pt}{3pt}
        \textbf{3} & Indicates that compared with two factors, \\ &  one factor is slightly more important than the other \\ 
        \specialrule{0em}{3pt}{3pt}
        \textbf{5} & Indicates that when two factors are compared,  \\ & one factor is significantly more important than the other \\ 
        \specialrule{0em}{3pt}{3pt}
        \textbf{7} & indicates that when two factors are compared, \\ &  one factor is stronger than the
other \\ 
        \specialrule{0em}{3pt}{3pt}
        \textbf{9} & Indicates that when two factors are compared, \\ &  one factor is extremely important than the other \\ 
        \specialrule{0em}{3pt}{3pt}
        \textbf{2,4,6,8} & The median value of the above two adjacent judgments \\
        \specialrule{0em}{3pt}{1pt}\hline
    \end{tabular}
    
\end{table}

Combining  the  previous  analysis  with  the  existing experience, the comparison matrix is shown in Table \uppercase\expandafter{\romannumeral3}.
\begin{table}[!ht]
    \centering
    \caption{Factors Comparison Matrix}
    \begin{tabular}{ccccc}
    \hline
        ~ & \textbf{NoR} & \textbf{TC} & \textbf{NoS} & \textbf{Cost}  \\ \hline
        \textbf{NoR} & 1 & 0.333 & 22 & 8  \\ 
        \textbf{TC} & 3 & 1 & 5 & 6.024 \\ 
        \textbf{NoS} & 0.5 & 0.2 & 1 & 8   \\ 
        \textbf{Cost} &  0.125 & 0.166 & 0.125 & 1  \\ \hline
    \end{tabular}
\end{table}

\subsubsection{Hierarchical single ranking to obtain weight and its consistency check}
The eigenvector corresponding to the largest eigenroot $\lambda$ max   of  the  judgment   matrix   is   denoted   as   W   after normalization (so that the elements in the vector sum to one). The  element  of W  is  the  ranking  weight  of the  relative importance of the element of the same level to the factor of the upper level. This process is called hierarchical single ranking.
Defining consistency measures:
\begin{equation}
\mathrm{CI}=\frac{\lambda-\mathrm{n}}{\mathrm{n}-1}
\end{equation}

\begin{itemize}
    \item CI=0, complete consistency
\item CI was close to 0, indicating satisfactory consistency
\item The larger the CI, the more severe the inconsistency

\end{itemize}

To measure the magnitude of CI, the random agreement index RI is introduced, which is shown in Table \uppercase\expandafter{\romannumeral4}.
\begin{table}[!ht]
    \centering
    \caption{Index of Consistency}
    \begin{tabular}{cccccccccccc}
    \hline
        \textbf{n} & 1 & 2 & 3 & 4 & 5 & 6 & 7 & 8 & 9 & 10  \\ \hline
        \textbf{RI} & 0 & 0 & 0.58 & 0.90 & 1.12 & 1.24 & 1.32 & 1.41 & 1.45 & 1.49  \\ \hline
    \end{tabular}
\end{table}

The consistency ratio was defined as CR = CI/RI. When the ratio is less than 0.1, the degree of inconsistency of A was considered to be within the allowable range, and there was satisfactory consistency, which passed the consistency test. Its normalized  eigenvector  can be used  as the weight vector, otherwise,   the   pairwise   comparison   matrix   A   must   be reconstructed.

After calculation, the parameters of the contrast matrix are as follows:
\begin{itemize}
    \item The largest characteristic root $\lambda$ :4.396
\item Vector of features: [1.52, 3.083, 0.946, 0.226]
\item Value of weight: [26.318, 53.395, 16.379, 3.098]
\item CI: 0.102 
\item RI: 1.12
\item CR: 0.0911 $\textless$ 0.1

\end{itemize}

Therefore, it passes the consistency test.

In  the  end,  we  obtained  mathematical  models  with weights  related  to  the  number  of residents,  transportation convenience,  the  number  of health  care  workers,  vaccine storage, and transportation costs, which can be used to make predictions of vaccine allocation proportions by multiplying these weights with characteristic variables in different regions.
\section{Model and Solution of Problem 3}
We mainly relied on the evaluation model established in question 2 to analyze the number or proportion of vaccines distributed  by  central  hospitals,  community  hospitals  and health  centers  in  Gongshu  district,  Hangzhou  and  Daoli district,   Harbin. The maps of Gongshu  district and Daoli district are shown in Fig. \ref{fig:6} and Fig. \ref{fig:7}. In   addition,   taking   into   account   the differences in basic conditions such as street communities and traffic in the two districts, we fine-tuned the model and fully cooperated with the individual characteristics of each district.

We collected the number of nearby permanent residents, streets,  communities,  medical  personnel  and  medical  and health institutions in Gongshu District, Hangzhou City and Daoli  District,  Harbin  City,  and  evaluated  the  distance between  the  central hospital,  community hospital  and the Manhattan residents' gathering center and the distribution of health centers through the map, and got the evaluation value of the degree of transportation convenience. At the same time, according to the economic conditions of the two places and the actual construction of medical institutions at all levels in the public data of the government, the storage cost indicators of central hospitals, community hospitals and health centers were obtained. The data is shown in Table \uppercase\expandafter{\romannumeral5} and Table \uppercase\expandafter{\romannumeral6}.

\begin{table}[!ht]
    \centering
    \caption{Gongshu District, Hangzhou}
    \begin{tabular}{ccccc}
    \hline
        \textbf{Gongshu District}  & \textbf{NoR} & \textbf{TC} & \textbf{NoS} & \textbf{Cost} \\ \hline
        \textbf{CenH} & 2.041 & 0.3 & 2.049 & 0.3  \\ 
        \textbf{ComH} & 2.014 & 0.5 & 1.729 & 0.2  \\ 
        \textbf{HC} & 1.513 & 0.6 & 0.853 & 0.1 \\ \hline
    \end{tabular}
\end{table}

\begin{table}[!ht]
    \centering
    \caption{Daoli  District,  Harbin }
    \begin{tabular}{cccccc}
    \hline
        \textbf{Daoli  District} & \textbf{NoR} & \textbf{TC} & \textbf{NoS} & \textbf{Cost} \\ \hline
        \textbf{CenH} & 2.040 & 0.3 & 2.545 & 0.9   \\ 
        \textbf{ComH} & 1.938 & 0.6 & 0.992 & 0.6   \\ 
        \textbf{HC} & 1.510 & 0.8 & 0.437 & 0.1 \\ \hline
    \end{tabular}
\end{table}

\begin{figure}[h]
    \centering
    \includegraphics[width=0.48\textwidth]{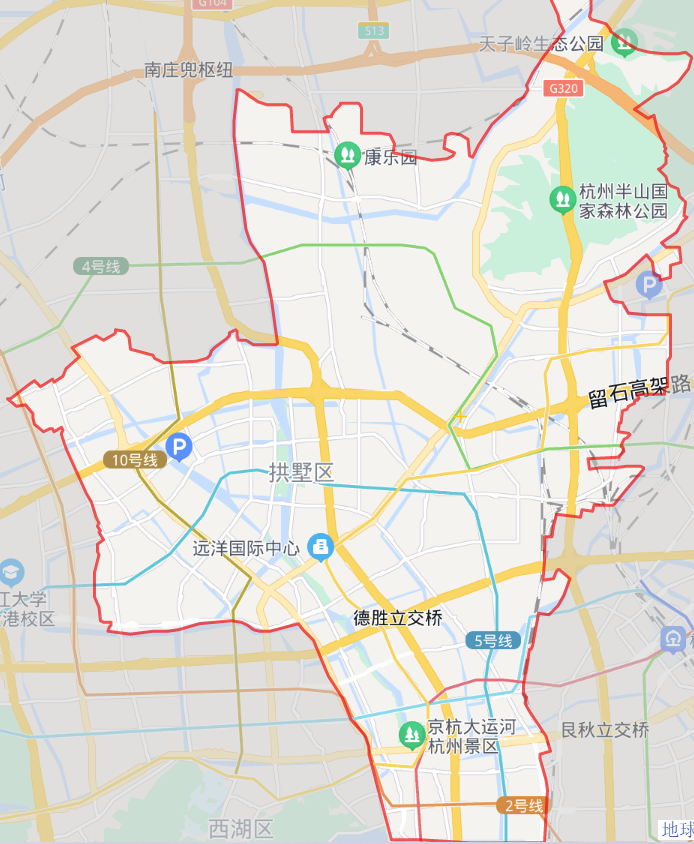}
    \caption{The map of Gongshu district}
    \label{fig:6}
\end{figure}

\begin{figure}[h]
    \centering
    \includegraphics[width=0.5\textwidth]{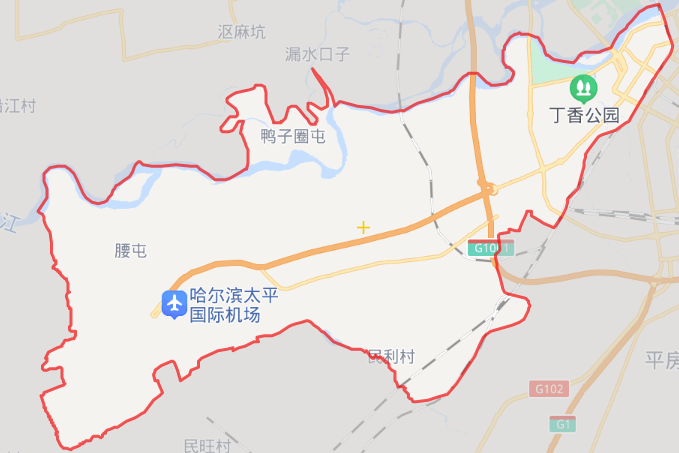}
    \caption{The map of Daoli district}
    \label{fig:7}
\end{figure}

Substitute the weight in question 2 to solve the vaccine distribution index of central hospitals, community hospitals and  health  centers. And subtract 0.1 * number  of  nearby residents from the index of central hospitals as a penalty for crowd  gathering. Finally,  for  the  convenience  of practical application, the ratio of vaccine distribution index of central hospitals,  community  hospitals  and  health  centers  was calculated by an approximate fraction of one decimal place:

Hangzhou Gongshu District: central hospitals, community hospitals, health centers= 4.8:3.3:1.9.

Harbin  Daoli  District:  central  hospitals,  community hospitals, health centers= 3.6:4.7:1.7.
\section{Model and Solution of Problem 4}
\subsection{Vaccine Allocation Should First Follow an Optimized Scientific Vaccination Strategy}
\begin{itemize}
    \item Research background: According to recent studies, it is important that priority groups receive the vaccine  first before it is phased out to all populations. And within priority groups  of the  same  level, increasing  coverage  of basic vaccines has a greater impact on reducing the severity of social  illness  than  increasing  coverage  of booster  shots. While   in   different   priority   groups,   increasing   booster coverage in higher priority groups generally reduced social severity more than increasing basic vaccination coverage in lower priority groups. 
    \item Vaccination in China: The WHO's strategy to achieve   the   global   Covid-19   vaccination   target   by   mid-2022   considers that the vaccination coverage rate of more than 70\% is  a  very  high  level,  and  herd  immune  shielding  can  be   established. According to official data, till July 23, 2022,   The cumulative coverage rate of the SARS-CoV-2 vaccine was   89.7\%, and the booster vaccination rate was 71.7\%. Among   the elderly over 60 years old, the full vaccination rate was   84.7\%, and the booster vaccination rate was 67.3\%.  It can   be seen that China is now a country with a medium to high coverage rate of the original series of vaccines in the high-priority groups, and there is a balance between the key groups   and non-key groups, the original series of vaccines and the   booster vaccine. Therefore, to a certain extent, it has played a   positive effect on the use of vaccines.
    \item Optimization  of  distribution  strategy:  In  order  to achieve better vaccine immunization effect, before providing the vaccine to the lower priority use groups, high enhanced needle  coverage  among  higher  priority  use  groups:  the highest  priority  population  should  include  older  people, health workers, especially front-line workers in health care and social care environments related to COVID- 19 patients, patients with immunodeficiency; High priority groups that also  need  to  be  paid  attention  to  include  adults  with comorbidities,  pregnant  women  with  the severe  disease  after infection and considering specific vaccination, teachers and other basic staff who play an important role in maintaining normal social functioning, and other vulnerable social groups at higher risk of illness. Besides, Some humanitarian buffer can be retained: reserve vaccines for vulnerable groups such as   refugees,    asylum    seekers,    and   workers    in   this environment.

\end{itemize}
\subsection{Vaccine Distribution Needs to Ensure Flexibility and
Legitimacy on the Basis of Regional Differences
}
\begin{itemize}
    \item{Flexibility}: In addition to the population priorities by age  and  existing  physical  condition,  each  region  should develop  differentiated  vaccination  strategies  according  to their  actual  characteristics:  it  can improve the priority  of people with strong regional mobility and relatively closed living environment and  strengthen the priority of vaccine supply to areas with the serious epidemic situation. In addition, with relatively sufficient vaccines, the vaccination ofnon-key groups can be promoted as soon as possible when the vaccine coverage in key groups has reached a high level.
    \item{Legality}: Through good and effective communication and publicity, the optimized vaccine distribution strategy can achieve the legitimacy principle in the values framework. Relevant governments can enhance the active participation of stakeholders in building common values through transparent and  open  scientific  data  and  expertise,  and  delegate  the agreed   results,   promote   community   participation   in implementation, and build multi-party trust.
\end{itemize}
\section{Strength and Weakness}
\subsection{Strength}
\begin{itemize}
    \item With  regard  to  vaccine   allocation,  we  give   full consideration to various theoretical and practical issues, make our thinking as close to the actual situation as possible, and use the setting of different indicators and weights to reflect our thinking in the model.
\item For the two problems of prediction and comprehensive evaluation, we adopted a model that is recognized as having an excellent effect in the field: LSTM and AHP.
\item Our team members successfully collected various data through various information channels, which provided strong realistic support for the establishment of our model.

\end{itemize}
\subsection{Weakness}
\begin{itemize}
    \item The robustness and sensitivity of the model are not discussed deeply, which is left for further research.
\item More mathematical models can be used for fitting to fully explore the rules of vaccine distribution.
\item Data of only two provinces have been collected, and more data are needed to improve the model.

\end{itemize}
\section{Conclusion}
We used machine learning methods and mathematical models to predict the number of future vaccinations and solved the problem of how to distribute vaccines to central hospitals, community hospitals and health centers. The time series prediction method LSTM was used to predict the next three months, and AHP was used to analyze the impact of the number   of  nearby  residents,   convenient   transportation, number   of   medical   personnel,   vaccine    storage   and transportation  cost  on  vaccine  distribution. And  received central hospitals, community hospitals, and health centers vaccine distribution ratio: Hangzhou Gongshu District with 4.8:3.3:1.9; Harbin Daoli District  with 3.6:4.7:1.7. Finally, combined with our model and conclusion, the vaccine distribution is fully explained. Our   model   provides   a   comprehensive   and advanced vaccine distribution  scheme, which can provide a useful reference for the existing distribution work. We hope to continue to improve our model in the future, collect more data, and strengthen the universality and robustness of the model.

\bibliographystyle{IEEEtran}
\bibliography{IEEEabrv, IEEEexample,my,my2}


\end{document}